\documentstyle[prd,floats, aps]{revtex}
\input psfig
\pssilent
\begin{document}

\title{Close limit of grazing black hole collisions: \\
non-spinning holes}

\author{Gaurav Khanna$^1$, Reinaldo Gleiser$^2$, Richard Price$^3$,
Jorge Pullin$^1$}

\address{1. Center for Gravitational Physics and Geometry,
Department of Physics\\
The Pennsylvania State University, 104 Davey Lab, University Park PA 16802}
\address{2. Facultad de Matem\'atica, Astronom\'{\i}a y F\'{\i}sica,
Universidad Nacional de C\'ordoba,\\
Ciudad Universitaria, 5000 Cordoba, Argentina}
\address{3. Department of Physics, University of Utah, Salt Lake City, UT
84112}
\date{November 16th 1999}
\maketitle

\begin{abstract}
Using approximate techniques we study the final moments of the 
collision of two (individually non-spinnning) black holes which
inspiral into each other. The approximation is based on treating the
whole space-time as a single distorted black hole. We obtain estimates
for the radiated energy, angular momentum and waveforms for the
gravitational waves produced in such a collision. The results can be
of interest for analyzing the data that will be forthcoming from
gravitational wave interferometric detectors, like the LIGO, GEO, LISA,
VIRGO and TAMA projects.
\end{abstract}

\section{Introduction}

The phrase ``collision of black holes'' has an
aura of a mysterious and exotic happening that is not far from the
reality of such an event. A black hole is not an ordinary object
defined by the amount and properties of the material of which it is
made. Rather it is a region from which no signal can escape. The
surface, the black hole horizon, bounding this region is defined by
the formal ``no escape'' property. Unlike the surface of an ordinary
object the horizon has no local properties that would be sensed by an
observer with the bad fortune to fall inward through it. A collision
of two holes is the process in which two no-escape regions merge
to become a single, larger, region of no escape.  In the last few
years such mergers have become the focus of much research attention,
for two not entirely independent reasons.

The first reason is the development of numerical relativity
\cite{Segr}. General relativity, Einstein's theory of gravity, 
sets the dynamics of space-time via 
a set of nonlinear partial differential equations of such complexity
that analytic solutions have been limited to two classes: solutions of
high symmetry, or solutions based on approximation techniques, such as
linearized weak field theory. The study of Einstein's equations on
computers has been viewed as the key to finding more general
asymmetric strong field solutions and it was natural for this key to
be be applied to black hole collisions. Black holes are
incontrovertibly strong field regions, but single isolated black holes are
stationary solutions of Einstein's theory, and the simplifying
symmetry of time independence allows for closed form well-understood
solutions \cite{Sch}. Collisions of black holes, on the other hand, are
necessarily nonstationary as well as being crucially strong-field
events.  It is known that the collision will result in a single final
black hole and in the generation of gravitational waves carrying off
some of the mass energy originally associated with the holes. But this
is all that is known with certainty. The nature of the merging of the
horizons, in the general collision, is not even qualitatively
understood.

A reasonably complete understanding awaits progress in numerical
relativity, and the wait has been longer than anticipated. The
solution of general black hole collisions on computers has proved to
be remarkably difficult. There is, however, a class of cases in which
reliable answers are available. If the collision is a 'head-on'
collision along a straight line, then there is rotational symmetry
about the line of the collision. Though the collision is still highly
dynamic and nonlinear, the simplifications afforded by this symmetry
reduce the computational demands sufficiently that the collision could
successfully be simulated even in the mid 1970's, and run with good
reliability in the mid 1990's\cite{NCSA}. The simplification of
head-on collisions, however, masks some of the physics of the most
interesting types of collisions, the fully three dimensional
collisions at the end point of the inspiral of a mutually orbiting
pair of black holes.

The second development that directed attention to black hole
collisions is the advent of sensitive gravitational wave detectors.
In the next few years, several interferometric gravitational wave
observatories (the LIGO project in the US, the VIRGO and GEO projects
in Europe and the TAMA project in Japan \cite{gw}) may be capable of
detecting gravitational waves. Whether near term searches are
successful will depend more than anything else on the strength of
astrophysical sources. Attributes of a good generator of gravitational
waves include strong gravitational fields and high velocities, so
black hole processes are a natural source to consider. It is
astrophysically plausible that black holes form binary associations
with other objects, including other black holes \cite{nyt}. Due to the
loss of energy by the emission of gravitational radiation, the
separation and period of the binary orbits would decrease. If the
binary consists of two black holes, the inspiral would end with a
rapid strong field merger that has the potential to be a powerful
source of detectable gravitational waves \cite{Tho}.

The whole process of inspiral generates gravitational radiation, but
in the early large-separation stages the radiation is relatively weak
and is reasonably well described by Newtonian gravity theory and
Post-Newtonian extensions of it \cite{Cliff}.  It is only the final
strong field merger that could in principle produce a powerful burst
of gravitational waves, but at this point only one parameter of the
burst is reliably known.  The characteristic frequency of the waves is
inversely proportional to the mass of the final black hole formed, and
works out to be on the order of $10^{3}$\,Hz for a 10$M_{\odot}$ hole,
a typical expected mass of a ``stellar'' sized hole. For supermassive
holes of mass $\geq 10^{6}M_{\odot}$ typical of galactic nuclei, the
waves would be less than 1\,Hz. The maximum sensitivity of the next
generation of gravitational wave detectors occurs at frequency around
100\,Hz and the detectors will be ideally suited to waves from a
black hole with mass of several hundred $M_{\odot}$. Some
recent observations\cite{midweight} offer indirect evidence that black
holes in this range may exist.  If they do not, then the detection of
the collision of black holes may require the deployment of space-based
detectors\cite{LISA} sensitive to the low frequency waves produced by
supermassive holes.

The ratio of the masses in a binary determines both how difficult it
is to analyze, and how exciting it is as a potential source. If the
mass of a black hole $M_{1}$ is much larger than the mass of its
binary companion $M_{2}$, then the smaller mass object can be treated
as a perturbation to the well understood spacetime of the larger mass
black hole. The equations that describe perturbations are linear, and
hence relatively easily dealt with in general. In the specific case of
perturbations to black hole spacetimes, the techniques of calculation
were worked out in the 1970s and resulted in the 
Regge-Wheeler and Zerilli equations \cite{ReWh,Z} for perturbations of
Schwarzschild (nonrotating) black holes, and in the Teukolsky
\cite{Te} equation for perturbations of Kerr (rotating) black holes.
The relatively easily analyzed \cite{dripp} ``particle limit'' case
$M_{2}\ll M_{1}$ may be of interest in connection, say, with neutron
stars merging with supermassive black holes, but this process cannot
give the hoped for high power. It is easy to show the gravitational
wave power generated scales in the masses as $(M_{2}/M_{1})^{2}$. High
power requires roughly equal masses, and this means the
simplifications of the particle limit do not apply to the most
interesting sources.

If not directly applicable to equal mass inspiral, the clarity of the
particle limit can, at least, help us to formulate questions about the
nature of the endpoint of inspiral, like the existence of a last
stable circular orbit. As a particle orbits a black hole it reaches a
radius at which it can no longer stably orbit with slowly decreasing
radius and it begins a rapid inward plunge. For the inspiral of two
roughly equal mass holes it can be imagined that the binary gradually
spirals inward or that it reaches a point at which a discontinuous
plunge begins. If the late orbits are being degraded rapidly enough by
the emission of gravitational radiation, there might not even be any
meaning to late stage ``stability.''  This uncertainty about even the
qualitative nature of the late stage of the inspiral is related to an
important, but totally unresolved, question: How does the inspiraling
binary shed enough angular momentum to form a black hole?  In a
relativist's units in which $c=G=1$, a black hole must have a total
angular momentum $J$ that is limited by the maximum angular momentum
$J= M^{2}$ that a rotating (Kerr) black hole can possess. Until the
binary pair is close, its angular momentum will be above this limit,
but technical considerations\cite{angmom} limit the rate at which
angular momentum can be shed in gravitational waves at very late
stages. If both black holes of the pair are rapidly rotating with
angular momentum in the same direction the shedding appears to present
a barrier to the formation of the final single black hole. It is
possible that even the qualitative details of the late stage inspiral
depend on the angular momentum of the inspiraling binary.

The set of possibilities is considerable and  the answers are important
both to our understanding of nonlinear gravitational interactions and
to an understanding of gravitational wave sources. Real answers will
require advances in numerical relativity that will be several years in
coming, but interest in the questions justifies approximation methods
that can help, even slightly, to close some of the wide open
questions. We take such an approach here. We offer an estimate of the
gravitational radiation generated during the late stage of inspiral of
two black holes. Our method involves a number of assumptions and
limitations that constrain its applicability and reliability, but for all
its shortcomings it
is one step towards a complete understanding.

The approximation method we use, the ``close limit,'' \cite{Pujap} 
takes advantage
of the property of a black hole horizon. Late in the merger of the
binary the single horizon of the final black hole engulfs the entire
binary. All the complex structure of the binary will be inside that
final horizon, and cannot influence spacetime outside the horizon. It
is only what is outside the horizon that can generate gravitational
waves that can be detected by distant observers.  Since the ultimate
fate of the merger is a stationary black hole, it follows that
sufficiently late in the merger what is outside the hole will be a
perturbation of the final stationary hole. Thus the gravitational
waves generated during the latest stage of inspiral can be computed
using the techniques of perturbations of black hole spacetimes, with
the Zerilli, Regge-Wheeler, and Teukolsky equations.

To understand how the close limit method is to be used, it is
necessary to consider the general problem addressed by numerical
relativity. Einstein's field equations are divided into ``initial
value'' equations and equations of time evolution \cite{York}. The
initial value equations determine the nature of spacetime at a chosen
initial moment. The solutions of these initial value equations are the
initial values for the remaining differential equations of Einstein's
theory, the equations that determine the spacetime (including its
gravitational wave content) to the future of the initial time.  The
two tasks of numerical relativity are first to find an initial value
solution representing a moment in the life of the colliding holes, and
second to find the future spacetime for those initial values.
The more computationally difficult task is that of evolving to the
future and the codes that accomplish this task tend to be unstable for
long time evolutions.  For long evolutions to be avoided, the initial
value solutions must be chosen to be a moment late in the life of the
inspiral. If that moment is late enough, the close limit method can be
brought to bear and evolution can be carried out with the stable
linearized equations of perturbation theory. But choosing too late a
starting moment for evolution creates a new difficulty.

The connection of an initial value solution to a ``sensible'' physical
configuration for the binary is reasonably secure only if the binary
pair is well separated. At close separations, the gravitational field
of each of the binary holes strongly affects the other hole, and the
individual mass, individual angular momentum, and physical separation
of the holes lose clear meaning. The problem then requires navigating
between the Scylla of numerical instabilities for evolution, and the
Charybdis of uncertain initial conditions. By using a very late
initial moment and linearized evolution, the close limit method
completely avoids the former hazard.

There are reasons beyond speculation to believe that close limit
evolutions give useful answers.  Numerical relativity results are
available for axisymmetric head-on collisions\cite{NCSA}. These
represent evolution of a number of initial value solutions, in
particular the closed form solution due to Misner\cite{Mi}, containing
a single parameter representing the initial separation of equal mass
holes in units of the mass of the spacetime (in $c=G=1$ units).  This
separation index defines a parameterized family of initial value
solutions.  Choices of this parameter can be made corresponding to
large or small initial separation. When numerical relativity and close
limit results are compared it is seen that agreement is excellent for
small initial separations, and is surprisingly good even when the
initial configuration is not close enough for a horizon to engulf the
entire binary\cite{PrPu,etal}.  Arguments can be made also, that the
gravitational waves calculated in the late stage of inspiral are not
highly sensitive to details of initial data. Particularly interesting
in this regard is work by Abrahams and Cook\cite{AbCook}.

In the past several years the close limit method has been extensively
studied for head-on collisions of boosted and spinning holes 
and compared with the results of
numerical relativity.  Most notably, second order perturbation theory
has been developed for the close limit method\cite{2nd}. In this
process of comparison much has been learned about the strengths and
limitations of the close limit method, with the goal of applying the
method to problems that cannot yet be handled with numerical
relativity.  The present work represents the first example of this. We
report here the results of the application of the close limit method
for the three dimensional problem of the late stage inspiral of two
black holes.

We will use the close limit method for the initial data families
constructed by the Bowen and York \cite{BoYo} method and the
associated ``punctures'' families \cite{BB}. It is known that these
families possess an artificial radiation content when one considers
black holes that are close, but such content is also known to be
moderate \cite{GlNiPrPuspin}. An important advantage of these methods
is that they are typically the starting point for numerical
relativity, and thus close limit evolution of these starting points
can be compared with the numerical evolution of these same initial
data when such evolutions become available. The most important
disadvantage, for our purposes, is that the Bowen-York family does not
include the Kerr solution, the solution for a rotating hole. This
precludes finding a family of initial value solutions that goes, in
the limit of small initial separation, to a Kerr black hole. With
Bowen-York initial value solutions, then, we cannot consider a
collision that will result in a rapidly rotating hole. Rather, we
limit our attention to collisions involving a modest amount of total
angular momentum and consider the angular momentum as well as the
initial separation to be a perturbation of a nonrotating final hole.
It it should also be mentioned that currently fashionable
astrophysical scenarios suggest that the individual holes might not
carry a significant amount of spin \cite{nyt} in realistic black hole
collisions.

The organization of this paper is as follows: in the next section we
review the method for obtaining the initial data and describe the 
approximations involved. In the following two sections we discuss how to
set  up the perturbative formalism geared towards evolution. Since the
collisions have net angular momentum we will evolve them {\em both} as
a perturbation of a rotating and a non-rotating black hole. The
comparison of both approaches is given in the subsequent section and we
will see that insight is gained by treating the problem in two
different ways. We end with a discussion of the results in terms of
waveforms and  radiated energies and we describe a puzzle in the
calculation of the angular momentum radiated. 

For the reader who wishes to be spared all the details, we summarize
our results in a brief punchline: the final ringdown of the
inspiraling collision of two non-spinning black holes is unlikely to
radiate more than $1\%$ of the mass of the system or more than $0.1\%$
of its angular momentum in gravitational waves.

\section{Initial data}

To evolve a spacetime in general relativity, one needs to provide
initial data, a 3-geometry $g_{ab}$ and an extrinsic curvature
$K_{ab}$, that solve Einstein's equations on some starting
hypersurface (i.e., at some starting time). For two black holes, this
is an easy task if the holes are far apart, since one can superpose
the solutions for two individual holes ignoring their
interactions. When the black holes are close on the initial
hypersurface, the astrophysically correct initial data is the solution
corresponding to what would have evolved during the binary inspiral,
but such an evolution cannot be computed with present day techniques.
One must therefore use a somewhat artificial initial data solution
that is a best guess at a representation of close black holes.  The
need for such a guess is one of the sources of uncertainty in our
result.

\subsection{Summary of the Bowen--York construction:} 
\label{BoYosec}
The initial value equations for general relativity are,
\begin{eqnarray}
\nabla^a (K_{ab} - g_{ab} K) &=& 0\\
{}^3R-K_{ab} K^{ab} + K^2 &=&0
\end{eqnarray}
where $g_{ab}$ is the spatial metric, $K_{ab}$ is the extrinsic
curvature and ${}^3R$ is the scalar curvature of the three metric. If
we propose a 3-metric that is conformally flat $g_{ab} = \phi^4
\widehat{g}_{ab}$, with $\widehat{g}_{ab}$ a flat metric, and $\phi^4$
the conformal factor, and we use a decomposition of the extrinsic
curvature $K_{ab} = \phi^{-2} \widehat{K}_{ab}$, and assume maximal
slicing $K_a^a=0$, the constraints become,
\begin{eqnarray}
\widehat{\nabla}^a \widehat{K}_{ab} &=& 0\label{momentum}\\
\widehat{\nabla}^2 \phi &=& -\frac{1}{8}
\phi^{-7} \widehat{K}_{ab} \widehat{K}^{ab}\ ,\label{hamihami}
\end{eqnarray}
where $\widehat{\nabla}$ is a flat-space covariant derivative.

To solve the momentum constraint, we start with a solution that 
represents a single hole with linear momentum $P$ \cite{CoYo},
\begin{equation}
\hat{K}^{\rm one}_{ab} = {3 \over  2 r^2} \left[ 2 P_{(a} n_{b)} 
-(\delta_{ab}-n_a n_b)P^c n_c\right]\ .
\end{equation}
In this expression for the conformally related extrinsic curvature at
some point $x^a$, the quantity $n_b$ is a unit vector, in the ``base''
flat space with metric $\hat{g}_{ab}$, directed from a point
representing the location of the hole to the point $x^a$. The symbol
$r$ represents the distance, in the flat base space, from the point
of the hole to $x^a$. It is straightforward to show that 
the solution of the Hamiltonian constraint corresponding to eq. (5)
corresponds to a spacetime with  ADM momentum $P_a$.

The next step is to modify this to represent holes centered at 
$x=\pm L/2$ in the conformally flat metric. Since the 
momentum constraint is linear, we can simply add two expressions
of the above form,
\begin{equation}
\widehat{K}^{\rm two}_{ab} =
\widehat{K}^{\rm one}_{ab}\left(x \rightarrow x-L/2,{P_y = P} \right)+
\widehat{K}^{\rm one}_{ab}\left(x \rightarrow x+
L/2,{P_y} = - P \right)\ .
\end{equation}
We will choose
in further expressions to use a polar coordinate system in the flat
space determined by $\hat{g}_{ab}$
centered in the mid-point separating the two holes and label the polar
coordinates as $(R,\theta,\phi)$. So $R$ will be the distance in the
flat space from the midpoint between the holes.

To solve the Hamiltonian constraint \ref{hamihami}, we introduce an
approximation, (the slow approximation) which we will show is enough
for our purposes. In fact, in this approximation the solution for the
conformal factor turns out to be the familiar Misner \cite{Mi}
solution if one chooses the topology of the slice to have a single
asymptotically flat region, or the Brill--Lindquist \cite{BrLi}
solution if there are three asymptotically flat regions.

\subsection{The slow approximation}

We assume that the black holes are initially close, and that the
initial momentum $P$ is small. We denote by $\vec{n}^+$ and
$\vec{n}^-$ the normal vectors corresponding, respectively, to the one
hole solutions at $x=+L/2$ and at $x=-L/2$, and we define $R$ to be
the distance to a field point, in the flat conformal space, from the
point midway between the holes.  For large $R$, the normal vectors
$\vec{n}^+$ and $\vec{n}^-$ almost cancel.  More specifically
$\vec{n}^+=-\vec{n}^-+O(L/R)$. A consequence of this is that the total
initial $\widehat{K}^{ab}$ is first order in $L/R$, and its
($R,\theta,\varphi$ coordinate basis) components can be written as
\begin{equation}
\hat{K}_{ab} = {3 P L \over 8R^3} \left[
\begin{array}{ccc} 8 \sin^2 \theta \sin 2\varphi&0&8 R \sin^2 \theta\\
0&R^2 (-5+\cos 2\theta) \sin 2\varphi&-2 R^2 \sin 2\varphi \sin 2\theta\\
8 R \sin^2 \theta&-2 R^2 \sin 2\varphi \sin 2\theta&R^2 \sin^2\theta 
\sin 2 \varphi (1+3 \cos 2\theta)
\end{array}\right]\ .
\end{equation}

This solution for $\hat{K}_{ab}$ is first order, both in $P$ and
$L$. Thus the source term in the Hamiltonian constraint is quadratic
in $P$. If we choose to find a solution to the conformal factor to
first order in $P$ (which should give us a good approximation in the
case of slowly moving holes), we can ignore this quadratic source
term. So now, the Hamiltonian constraint looks like the one for zero
momentum, which is simply the Laplace equation. A well known solution
to this, is the Misner solution \cite{Mi}. This solution, is
characterized by a parameter $\mu_0$ which describes the separation of
the two throats.  We can relate this parameter to the conformal
distance $L$ in the following way \cite{NCSA},
\begin{equation}
L/M=\frac{\rm{\coth}\mu_0}{2\Sigma_1}\ \ \hspace*{30pt}
  \ \ \ \Sigma_1\equiv\sum_{n=1}\frac{1}{\sinh n\mu_0}\ .
\end{equation}

To clarify: in the slow approximation we are considering, the data we
use in our simulations consists of the extrinsic curvature proposed by
Bowen and York and the conformal factor due to Misner. This might
appear as odd, since the conformal factor of Misner is ``symmetrized''
through the throats and the extrinsic curvature due to Bowen and York
is not. What we do is not inconsistent, it is just a different (and
perhaps from a certain point of view less natural) choice of boundary
conditions for the fields. In practice, in the close limit and to
first order in perturbation theory, the conformal factor of Misner
differs from that of Brill and Lindquist by a numerical factor that
can be absorbed in the definition of the separation of the holes
\cite{AbPr}.

Some readers may be disturbed by the slow approximation, since in the
computation of certain quantities, for instance the ADM mass, the
higher order terms in the expansion in terms of the momentum are
crucial. We have already discussed this in detail in previous head-on
simulations \cite{boost1}. The bottomline is that to get an
accurate estimate of the ADM mass for high values of the momentum one
indeed needs a full solution of the Hamiltonian constraint and not a
``slow approximation'' solution. For the values of the separations and
the momenta we will consider in this paper ($a<0.5$) the ADM mass
computed with the slow approximation and the one computed with the
full solution differ by less than $10\%$ so we will ignore this
difference.

We must now map the coordinates of the initial value solution to the
coordinates for the Schwarzschild/Kerr (in the vanishing spin limit)
background. To do this, we interpret the $R$  as the isotropic
radial coordinate of a Schwarzschild spacetime, and we relate it to
the usual Schwarzschild radial coordinate $r$ by
$R=(\sqrt{r}+\sqrt{r-2M})^2/4$.  From this we arrive at the following
expression for the components of the extrinsic curvature,
\begin{equation}
K_{ab} =
{3 P L \over 8r^3} \left[
\begin{array}{ccc} \frac{8 \sin^2 \theta \sin 2\varphi}{1-2M/r}&0&
\frac{8 r \sin^2 \theta}{\sqrt{1-2M/r}}\\
0&r^2 (-5+\cos 2\theta) \sin 2\varphi&-2 r^2 \sin 2\varphi \sin 2\theta\\
\frac{8 r \sin^2 \theta}{\sqrt{1-2M/r}} &-2 r^2 \sin 2\varphi \sin 2\theta
&r^2 \sin^2\theta \sin 2 \varphi (1+3 \cos 2\theta)
\end{array}\right]\ .
\end{equation}
Here we have used the fact that
\begin{equation}
\phi^2\approx\phi_{\rm Mis}^2\approx
\phi_{\rm Schw}^2=r/R
=\frac{1}{\sqrt{1-2M/r}}\,\frac{dr}{dR}\ .
\end{equation}

\section{The close limit as a perturbation of a Schwarzschild hole}

In this paper we will evolve the initial data we just constructed
using the perturbative evolution equations for linearized first order
perturbations: the Zerilli equation in the case of a Schwarzschild
background and the Teukolsky equation in the case of a Kerr
background. We need to construct the initial data for these equations
in terms of the metric and extrinsic curvature we discussed above.
In this section we discuss the setup of initial data and evolution 
of the problem as a perturbation of a Schwarzschild black hole, using
the Zerilli--Regge--Wheeler formalism.

\subsection{Setting up the initial data for the Zerilli function}

Given the three metric and the extrinsic curvature, one can explicitly
construct the zeroth and first order term of a power series expansion
in a fiducial time variable $t$ of the space-time metric. From this
expression one can read off the appropriate coefficients of the
multipolar expansion of the metric in the Regge--Wheeler \cite{ReWh}
notation. The only nonvanishing perturbations at $t=0$ are,
\begin{eqnarray}
H_2[\ell=2,m=\pm2] & =& K[\ell=2,m=\pm2] =  
\sqrt{{6 \pi \over  5}} {8 M L^2 \over
\sqrt{r} (\sqrt{r} + \sqrt{r-2M})^5} \nonumber \\
H_2[\ell=2,m=0] & = & K[\ell=2,m=0] =- 2\sqrt{{\pi \over 5}} {8 M L^2 \over \sqrt{r}
(\sqrt{r} + \sqrt{r-2M})^5}
\end{eqnarray}

We compute the time derivative of these quantities, using the 
extrinsic curvature $K_{ij}$ obtained in the last section. The
nonvanishing ones are,
\begin{eqnarray}
{\partial H_2[\ell=2, m=-2] \over \partial t} & = &  -i 24 \sqrt{{\pi
\over 30}} P L {\sqrt{r-2M} \over r^3 \sqrt{r}}\nonumber \\
{\partial H_2[\ell=2, m=2] \over \partial t} & = &  - {\partial
H_2[\ell=2, m=-2] \over \partial t}\nonumber \\
{\partial K[\ell=2, m=-2] \over \partial t} & = & i  \sqrt{30 \pi} P L
{\sqrt{r-2M} \over r^3 \sqrt{r}}  \nonumber \\
{\partial K[\ell=2, m=2] \over \partial t} & = &   -  {\partial
K[\ell=2, m=-2] \over \partial t} \nonumber \\
{\partial G[\ell=2, m=-2] \over \partial t} & = & i \sqrt{{6 \pi \over
5}} P L {\sqrt{r-2M} \over r^3 \sqrt{r}}
\nonumber \\
{\partial G[\ell=2, m=2] \over \partial t} & = &  - {\partial G[\ell=2,
m=-2] \over \partial t}
\end{eqnarray}
where $i$ is the imaginary unit. (Here we are using the standard
conventions for the spherical harmonics. Notice that the $m=2$ and
$m=-2$ perturbations are individually complex, but when they are added
to give the total perturbation the resulting function of $t,r,\theta$
and $\varphi$ is real, as of course it must be.

We also
have an odd parity contribution,
\begin{equation}
{\partial {\ }^{\mbox{{\small odd}}} h_1[\ell=1, m=0] \over \partial t}
= 8 \sqrt{3 \pi} {PL \over r^2}.
\end{equation}
This perturbation represents the difference between the Kerr solution
that represents the rotating space-time and the Schwarzschild
background used in the perturbative approach.  To first order it
decouples from all other perturbations, and in fact is unchanging in
time, corresponding to the conservation of angular momentum to first
order in the perturbations. The change over time in the quantity
induced by {\it second} order perturbations, will be discussed below
in connection with the radiation of angular momentum.

The Zerilli function is defined  by (see for instance \cite{GlNiPrPupr}),
\begin{eqnarray}
\psi_{(\ell,m)} & = &  {2r(r-2M) \over \ell(\ell+1)(\lambda r +3M)}\left[
 {{H}_2}_{(\ell,m)}
-r\frac{\partial   {K}_{(\ell,m)}}{\partial r}
-\frac{r-3M}{r-2M} {K}_{(\ell,m)}
\right] \nonumber \\
 \label{Mondef}
& & 
+{r^2 \over \lambda r +3M} \left[  {K}_{(\ell,m)}
+(r-2M)
\left(
 {\partial  {G}_{(\ell,m)} \over \partial r}
-{2 \over r^2}  {{h}_1}_{(\ell,m)}
\right) \right] .
\end{eqnarray}
 Therefore, for $t=0$ we have
\begin{equation}
\label{psi0}
\psi_{(2,m)}(0,r) =  {r(r-2M) \over 3(2r +3M)}\left[
 {{H}_2}_{(2,m)}(0,r)
-r {\partial   {K}_{(2,m)}(0,r) \over \partial r}\right]
+ {r \over 3} {K}_{(2,2)}(0,r).
\end{equation}
and
\begin{eqnarray}
\label{psidot0}
\dot{\psi}_{(2,m)}(0,r) & = & {r(r-2M) \over 3(2r +3M)}\left[
 { \dot{H_2}}_{(2,m)}(0,r)
-r {\partial   \dot{K}_{(2,m)}(0,r) \over \partial r} +3 r {\partial
\dot{G}_{(2,m)} \over \partial r} \right]
\nonumber \\
& & + {r \over 3} \dot{K}_{(2,2)}(0,r)  .
\end{eqnarray}

After some simplifications we have the initial data for the Zerilli
function,
\begin{equation}
\psi_{(2,2)}(0,r) = 4 \sqrt{{2 \pi \over 15}} M L^2 {r\;(7 \sqrt{r}+5
\sqrt{r-2M}) \over (2r+3M) (\sqrt{r}+\sqrt{r-2M})^5}
\end{equation}
and
\begin{equation}
\dot{\psi}_{(2,2)}(0,r)  =  -{ i \sqrt{30 \pi} \over 5} P L
{(4r+3M)\sqrt{r-2M} \over r^{5/2}(2r+3M)}
\end{equation}
and the Zerilli function for $(\ell=2,m=-2)$ is the complex conjugate
of $\psi_{(2,2)}(t,r)$. The initial data for the $\ell=2,m=0$ Zerilli function
is 
 \begin{eqnarray}
\psi_{(2,0)}(0,r) &=& -{8\over 3} \sqrt{\pi \over 5} {M L^2 r \; (7
\sqrt{r} + 5 \sqrt{r-2M}) \over (2r+3M) (\sqrt{r}+\sqrt{r-2M})^5}\\
\dot{\psi}_{(2,0)}(0,r)  &=&  0.
 \end{eqnarray}

\subsection{Evolution of the Zerilli function and computation of
physical quantities}

Given the Cauchy data from the last section, the time evolution is
obtained from the Zerilli equation \cite{GlNiPrPupr},
\begin{equation}
-{\partial^{2} \psi_{(\ell,m)} \over \partial t^{2}} +
{\partial^{2} \psi_{(\ell,m)} \over \partial r_{*}^{2}} +V(r_{*})
\psi_{(\ell,m)} 
= 0,
\end{equation}
where $V(r_{*})$ is the ($m$-independent) Zerilli potential,
\begin{equation}
V(r_*) = 2 \left(1 -{2 M \over r}\right) { \lambda^2 r^2
\left[(\lambda+1) r + 3 M \right] + 9 M^2 (\lambda r +M) \over r^3
(\lambda r + 3 M)^2 }\ ,
\end{equation}
where $\lambda = {(\ell-1)(\ell+2)}/{2}$ and $r_*=r+2M \ln(r/2M-1)$.

We need to establish a convenient formula for the radiated energy,
similar to that present in \cite{cpm} but applied to the
non-axisymmetric case. We start from the expression of the radiated
energy computed via the Landau--Lifshitz pseudo-tensor following the
notation and derivations of
\cite{cpm},
\begin{equation}
{d {\rm Power}\over d\Omega} = \lim_{r\rightarrow\infty} {1 \over 16\pi r^2}
\left[\left({\dot{h}_{\theta\phi}\over \sin\theta}\right)^2
+{1 \over 4} \left(\dot{h}_{\theta\theta}
-{1 \over \sin^2\theta} \dot{h}_{\phi\phi}\right)^2\right],
\end{equation}
and translating to the Regge--Wheeler notation and integrating on
solid angles we get,
\begin{equation}
{\rm Power}={3 \over 16\pi} \left[ 2 \left| \dot{\psi}_{(2,0)}\right|^2
+ 4 \left| \dot{\psi}_{(2,2)}\right|^2\right]
\end{equation}
and one can obtain the radiated energy integrating over time. The power
naturally comes out in units of the mass of the background spacetime.

To compute the radiated angular  momentum one could also start by
considering the Landau--Lifshitz pseudo-tensor and construct and
asymptotic expression for angular momentum flux. This approach was
pursued, for instance, in \cite{Tho80} to compute expressions for the
radiation of angular momentum in terms of multipoles. An alternative
approach is to simply compute the change in the angular momentum of
the spacetime, which we characterize to linear order in perturbation
theory through the function,
\begin{equation}
r^2 {\ }^{\mbox{\small{odd}}}h_0,_r(r,t) - 2 r
 {\ }^{\mbox{\small{odd}}}h_0(r,t) r^2
-{\ }^{\mbox{\small{odd}}}h_1,_t(r,t).
\end{equation}
This is a first order gauge 
invariant if ${\ }^{\mbox{\small{odd}}}h_0$, and 
${\ }^{\mbox{\small{odd}}}h_1 $
are first order perturbations.  Moreover, for $\ell=1, m=0$, this gauge
invariant is constant, equal to $4 \sqrt{3 \pi} J$, where $J$ is the
total angular momentum, if the perturbations are axially symmetric.

If we look at second order perturbations we find
\begin{equation}
{\partial \over \partial t} \left[r^2
{\ }^{\mbox{\small{odd}}}h_0,_r(r,t) - 2 r
{\ }^{\mbox{\small{odd}}}h_0(r,t) - r^2
{\ }^{\mbox{\small{odd}}}h_1,_t(r,t)\right] = {\cal{S}}_{\mbox{Jdot}}
\end{equation}
where ${\cal{S}}_{\mbox{Jdot}}$ is a `source', quadratic in first order
perturbations.

Therefore  the change in angular
momentum, due to radiation may be obtained by integrating
${\cal{S}}_{\mbox{Jdot}}$ for all $t$ (or from $t=0$ to $t = \infty$,
it makes no difference), in the limit $r \rightarrow \infty$. After 
several simplifications and cancelling terms that result from
integration by parts, we end up with
\begin{equation}
\Delta J = {3 i \over 4 \pi} \lim_{r \rightarrow \infty}
 \int_{0}^{\infty} \left[\psi_{(2,-2)}(r,t) {\partial
 \psi_{(2,2)}(r,t) \over \partial t} -\psi_{(2,2)}(r,t)
 {\partial \psi_{(2,-2)}(r,t) \over \partial t} \right] dt
 \end{equation}

If we write
\begin{equation}
\psi_{(2,2)}(r,t) = \mbox{Re}(\psi) + i \mbox{Im}(\psi) 
 \end{equation}
we have
\begin{equation}
\psi_{(2,2)}(r,t) = \mbox{Re}(\psi) - i \mbox{Im}(\psi) 
 \end{equation}
and we find
\begin{equation}
\Delta J = {3  \over 2 \pi} \lim_{r \rightarrow \infty}
 \int_{0}^{\infty} \left[ \mbox{Im}(\psi) {\partial \mbox{Re}(\psi)
 \over \partial t} - \mbox{Re}(\psi) {\partial \mbox{Im}(\psi) \over
 \partial t}\right] dt.
 \end{equation}
 
We have checked by explicit substitution that this form coincides with
the results from the flux formulas of Thorne
\cite{Tho80}. It reassures our confidence in the consistency of the
Regge--Wheeler--Zerilli perturbative formalism to notice that the
changes to second order are in accordance with the first order flux.

\section{Evolution as a perturbation of a Kerr black hole} 

To treat the problem as a perturbation of a Kerr black hole we need to
set up initial data and evolve the Teukolsky equation. The formalism
for setting up initial data in terms of Cauchy metric data was
developed in \cite{CaLoBaKhPu}, we only give a brief sketch here and
refer the reader to that paper for further details.

The relevant Weyl scalar for gravitational radiation is

\begin{equation}
\psi _4=-C_{\alpha \beta \gamma \delta }n^\alpha \overline{m}^\beta n^\gamma
\overline{m}^\delta ,
\end{equation}
since it is directly related to outgoing gravitational waves. We can
rewrite this as

\begin{equation}
-\psi _4=R_{ijkl}n^i\overline{m}^jn^k\overline{m}^l+4R_{0jkl}n^{[0}\overline{%
m}^{j]}n^k\overline{m}^l+4R_{0j0l}n^{[0}\overline{m}^{j]}n^{[0}\overline{m}%
^{l]},
\end{equation}
which in turn can be written in terms of hypersurface quantities
$g_{ij}$ and $K_{ij}$. For the last term in this expression, we can
use vacuum Einstein equations to eliminate terms that have time
derivatives of $K_{ij}$.  Also, we are interested merely in the first
order perturbations of this scalar. Putting all this together, the
final result for the first order expansion of the Weyl scalar is
\cite{CaLoBaKhPu},

\begin{eqnarray}
-\psi _4 &=&\left[ {R}_{ijkl}+2K_{i[k}K_{l]j}\right] _{(1)}n^i%
\overline{m}^jn^k\overline{m}^l-4N_{(0)}\left[ K_{j[k,l]}+{% \Gamma
}_{j[k}^pK_{l]p}\right] _{(1)}n^{[0}\overline{m}^{j]}n^k\overline{m}^l
\\ &&\ +4N_{(0)}^2\left[ {R}_{jl}-K_{jp}K_l^p+KK_{jl} \right]
_{(1)}n^{[0}\overline{m}^{j]}n^{[0}\overline{m}^{l]}
\end{eqnarray}
where $N_{(0)}=(g_{\text{kerr}}^{tt})^{-1/2}$ is the zeroth order
lapse, $ n^i,\overline{m}^j$ are two of the null vectors of the
(zeroth order) tetrad, Latin indices run from 1 to 3, and the brackets
are computed to only first order (zeroth order excluded).

This expression can be used, to obtain the time derivative of the
Weyl scalar too. We simply replace the first order quantities above by
their time derivatives (which can be obtained via the Einstein equations). 

In our treatment, the extrinsic curvature and the metric, from the
last section, shall be treated as a perturbation of the
corresponding Kerr hypersurface quantities. Since we attempt
calculations only to first order in $PL$ (which we identify with $Ma$,
where $M$ is the mass of the background Kerr black hole and $a$ its
angular momentum parameter), the Kerr 3-metric is (in this
approximation) conformally flat. Hence we justify using the Bowen York
recipe for constructing initial data for the inspiral problem.

\subsection{Initial Data for the Teukolsky function:}

Using the methodology and expressions we just discussed, 
the initial data for the Teukolsky function, $\Psi={\rho}^{-4}\psi_4$,
where ${\rho}={-1/{(r-ia\cos\theta)}}$, is:

For the azimuthal modes, $m=\pm 2$
\begin{equation}
-{\Psi\over\sqrt{2\pi}}= \left[{{3 r M(2M-r) L^{2}}\over{32
{R^2}(2R+M)}} \pm i {3\over 8} M a
\left(1-{2M\over r}\right)^{3\over 2}\right]{(\cos\theta \pm 1)^2}
\end{equation}
\begin{equation}
-{\dot\Psi\over\sqrt{2\pi}}=-\left[{{3 (2M-r) {M^2}
L^{2}}\over{16r{R^2}(2R+M)}}\pm i {3 M a\over 16 r^2} (2r - 21M) 
\left(1-{2M\over r}\right)^{3\over 2}\right]{(\cos\theta \pm 1)^2}
\end{equation}

And for the azimuthal mode, $m=0$
\begin{equation}
-{\Psi\over\sqrt{2\pi}}={{3 r M (r-2M) L^{2}}\over{16 {R^2}(2R+M)}}
\sin^{2}\theta 
\end{equation}
\begin{equation}
\dot\Psi = -{2M \over r^2} \Psi
\end{equation}
Here, $R$ is the Schwarzschild isotropic radial coordinate.

\subsection{Evolution of the Data using the Teukolsky equation}

Given the Cauchy data from the last section, the time evolution is
obtained from the Teukolsky equation \cite{Te},

\begin{eqnarray}
&&
\Biggr\{\left[a^2\sin^2\theta-\frac{(r^2 + a^2)^2}{\Delta}\right]
\partial_{tt}-
\frac{4 M a r}{\Delta}\partial_{t\varphi}
+ 4\left[r+ia\cos\theta-\frac{M(r^2-a^2)}{\Delta}\right]\partial_t
\nonumber\\
&&+\,\Delta^{2}\partial_r\left(\Delta^{-1}\partial_r\right)
+\frac{1}{\sin\theta}\partial_\theta\left(\sin\theta\partial_\theta\right)
+\left[\frac{1}{\sin^2\theta}-\frac{a^2}{\Delta}\right]
\partial_{\varphi\varphi}\\
&&-\, 4 \left[\frac{a (r-M)}{\Delta} + \frac{i \cos\theta}{\sin^2\theta}
\right] \partial_\varphi
-\left(4 \cot^2\theta +2 \right)\Biggr\}\Psi=0,
\end{eqnarray}
where $M$ is the mass of the black hole, $a$ its angular momentum per
unit mass, $\Sigma\equiv r^2+a^2\cos^2\theta$, and $\Delta\equiv
r^2-2Mr+a^2$.

The radiated energy is given by \cite{CaLo},

\begin{equation}
\frac{dE}{dt}=\lim_{r\to\infty}\left\{ \frac{1}{4\pi r^{6}}%
\int_{\Omega}d\Omega\left| \int_{-\infty}^{t}d\tilde{t}\ \Psi(\tilde{t%
},r,\theta,\varphi) \right|^2\right\}, \quad
d\Omega=\sin\theta\ d\vartheta\ d\varphi,
\end{equation}
and the angular momentum carried away by the waves can be
obtained from \cite{CaLo},
\begin{equation}
\frac{dJ_z}{dt}=-\lim_{r\to\infty}\left\{ \frac{1}{4\pi r^{6}}{\rm Re}\left[
\int_\Omega d\Omega
\left(\partial_\varphi\int_{-\infty}^{t}d\tilde{t}\
\Psi(\tilde{t},r,\theta,\varphi) \right)
\left(\int_{-\infty}^{t}dt^\prime\int_{-\infty}^{t^\prime}d\tilde{t}\ 
\overline{\Psi}(\tilde{t},r,\theta,\varphi)\right)\right]
\right\}.  
\end{equation}

\section{Results of the evolutions}

\begin{figure}[t]
\centerline{\psfig{figure=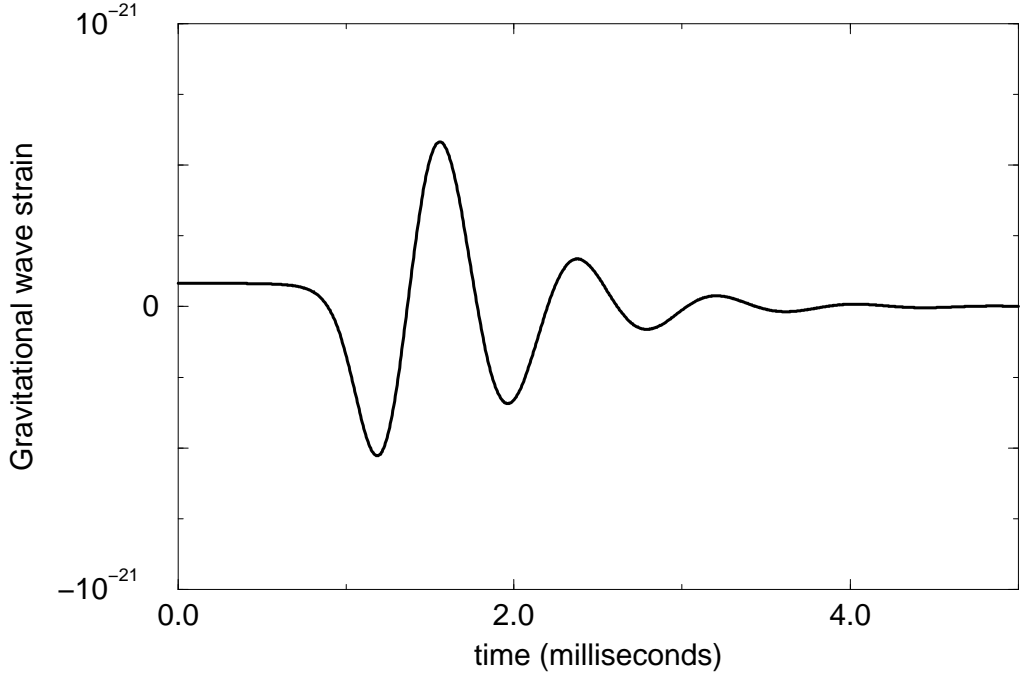,height=100mm}}
\caption{The gravitational wave one obtains from the close limit
non-head-on collision of two black holes. Depicted is the ``strain
amplitude'' of the ``+'' polarization mode 
in the equatorial plane (assuming the collision has
initial angular momentum aligned with the $z$ axis). We chose to
depict it in ``realistic'' units assuming that the binary has a mass
of $10M_{\odot}$ and we are observing the wave at a distance of
$100MPC$. The angle of observation is $\theta=\pi/2, \phi=0$. }
\label{fig1}
\end{figure}

We have evolved the Zerilli and Teukolsky equations using codes that
have already been tested in other situations \cite{boost1},
\cite{KrLaPaAn}. Figure \ref{fig1} shows the amplitude of the waves,
depicting the ``+'' component of the polarization, defined in terms of
the Zerilli function as,
\begin{equation}
h_+={1 \over r^2}\left(h_{\theta\theta}-{1 \over
\sin^2\theta}h_{\phi\phi}\right)= {\sqrt{30}\over 4\sqrt{\pi} r}
\left(\sqrt{6} \psi_{(2,0)}+(2-\sin^2\theta) \left[\psi_{(2,2)}
e^{2i\phi}+
\psi_{(2,-2)} e^{-2i\phi}\right]\right).
\end{equation}
\begin{figure}
\centerline{\psfig{figure=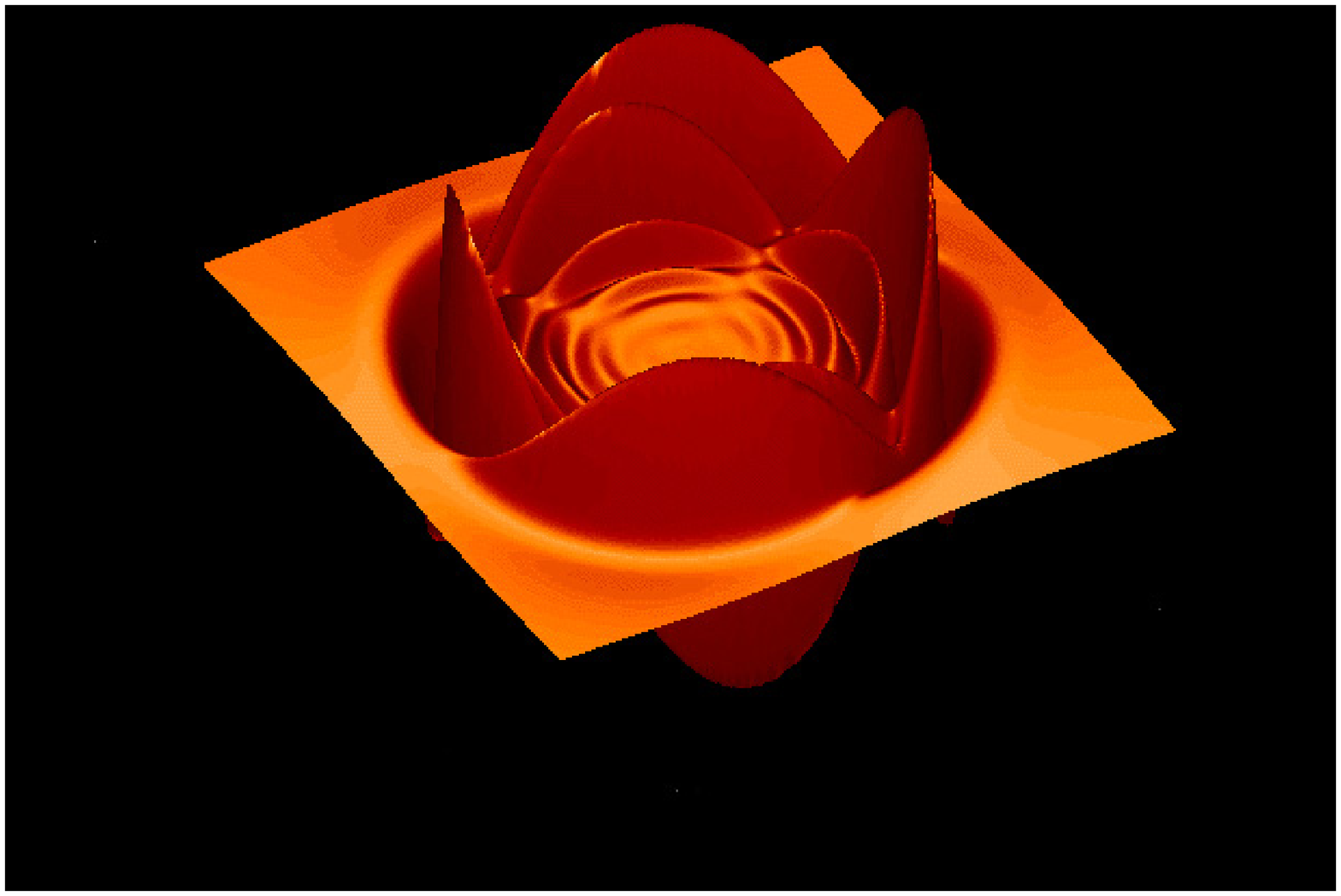,height=70mm}
\psfig{figure=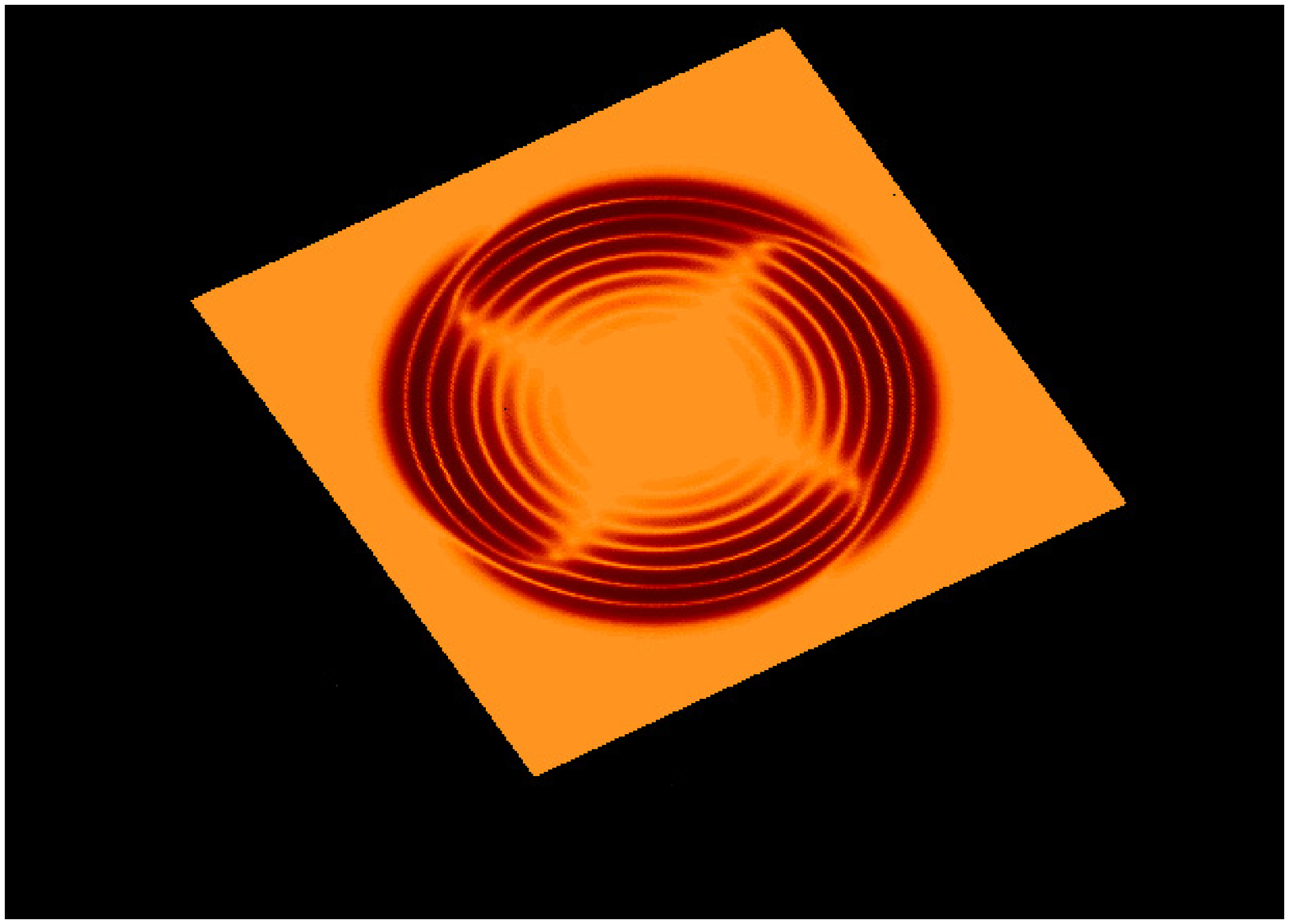,height=70mm}}
\caption{ A spatial rendition of the quantity: $ {i\sqrt{30}\over 8
\sqrt{\pi}}\left(\psi_{(2,2)} e^{2i\phi} -\psi_{(2,-2)}
e^{-2i\phi}\right)$. For large values of $r$ this is proportional to
the ``$\times$'' strain of the gravitational wave that a detector
would measure (see equation (24)). The proportionality factor is
$100Mpc/r$. The overall vertical scaling is the same as in the figure of the
quasi-normal ringing if $r=100Mpc$. In the attached movie one can see
the time evolution leading to these pictures. The viewer should keep
in mind that in order to visualize "strain" a factor of $100Mpc/r$
should be included, and the strain is only a well defined concept in
the far zone where the metric is approximately flat.  The plot is for
a given fixed colatitude $\theta=\pi/4$, as a function of $x,y$. The
left picture is a side view, the right picture a top view. The picture
corresponds to a snapshot at $t=80M$. The spatial scale is $\pm 100M$
in each direction. The vertical scale is the same as in figure 1. }
\label{fig2}
\end{figure}
In figure \ref{fig2} we give a spatial visualization of the waves, by
plotting the ``$\times$'' polarization of the waves, defined as,
\begin{equation}
h_\times={1 \over r^2} h_{\theta\phi} = {i\sqrt{30}\over 4 r
\sqrt{\pi}}
\cos\theta \sin\theta \left(\psi_{(2,2)} e^{2i\phi}
-\psi_{(2,-2)} e^{-2i\phi}\right),
\end{equation}
The figure suggests a
rotation pattern, but as can be seen in the accompanying movie, the
shown patterns just propagate outward.

\begin{figure}
\centerline{\psfig{figure=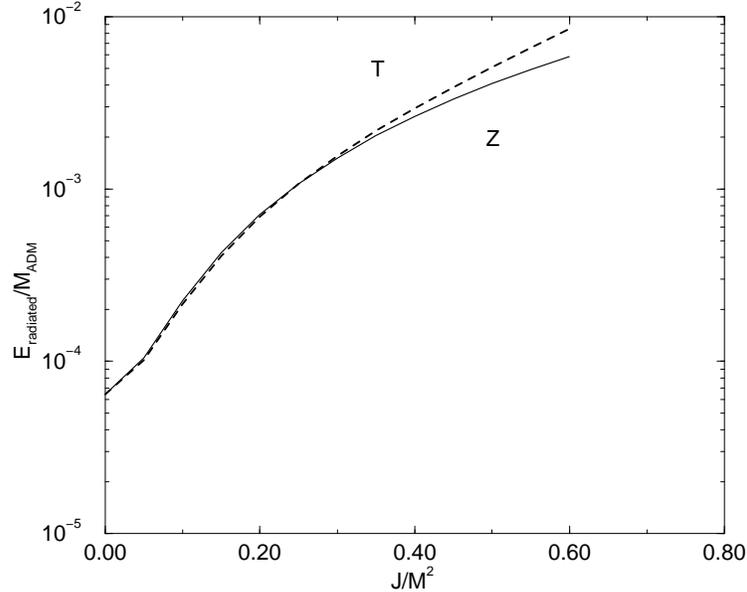,height=80mm}}
\caption{The radiated energy in a non-head-on collision of two
non-spinning black holes as a function of the total initial angular
momentum, for a fixed separation of $3.64$ radii (see text for
details). We depict the results of treating the problem as a
perturbation of a non-rotating hole (Z) and a rotating hole (T). The
agreement of both curves up to angular momenta of $a=0.4-0.5$ gives
confidence in the linear perturbative results.  The ``real data'' very
likely lies in a curve below the Zerilli (Z) curve, which allows us to 
roughly extrapolate the results to the extremal $a=1$ case, where we
see that still less than $1\%$ of the mass of the system is radiated
in the close limit.}
\label{fig4}
\end{figure}

Let us turn now to the evaluation of the radiated energies and angular
momentum. Figure \ref{fig4} shows the radiated energy as a function of
the initial angular momentum, for a fixed separation of the holes. The
figure compares the Regge--Wheeler--Zerilli (Z) and Teukolsky (T)
calculations. As expected, they differ for large values of the angular
momentum, since the Teukolsky calculation contains terms higher than
linear in the angular momentum. As we explained before, one is not
keeping consistently these higher order terms so one cannot argue that
the Teukolsky result is ``better''. A conservative view that can be
taken should be that both results disagree when higher order terms
start to be important, and this gives us a rough measure of the error
in the Zerilli calculation. We therefore conclude that for the
separation in question, one should not trust first order perturbation
theory beyond $a=0.5$. One should stress that this view can be
somewhat overconservative, our experience with explicit second order
calculations for the head-on collisions
\cite{GlNiPrPunsc,boost2,GlNiPrPupr} shows that one
should include {\em all} second order terms to have a consistent
formulation and a reliable set of ``error bars''. This is not
accomplished by the first order Teukolsky formalism in this
context. In this respect, second order Teukolsky results for this
problem will be quite welcome \cite{CaLo}. The second order Zerilli
calculations appear as quite prohibitive in complexity.
The separation of the holes quoted in figure \ref{fig4} requires some
explanation. The simulations start with the construction of the
initial data by the Bowen-York procedure we described in section
\ref{BoYosec}. As discussed there, the construction starts with the
introduction of a fiducial conformal space. In such a space the
separation is $0.91M$ where $M$ is the ADM mass of the spacetime. The
radius of each hole (if they were non-moving, the momentum slightly
changes the shape of the horizon and the radius, see \cite{CoYo}) is
approximately $M/4$, from there the separation of $3.64M$ quoted in
the caption. To translate to more commonly used terms, one could
convert the number to the $\mu_0$ parameter in the Misner solution,
which for our case is $\mu_0=1.5$. Finally, another commonly used
measure of the separation is the length of the geodesic threading the
throat in the Misner geometry. In terms of such a parameter, it is
equivalent to $2.75$ times the ADM mass of the spacetime or
approximately $5.5$ times the mass of each individual hole.

A remarkable aspect of figure \ref{fig4} is that linear perturbation
theory has a tendency to {\em overestimate} the radiated energies for
large values of the perturbative parameter, at least from our
experience with head-on collisions of non-boosted \cite{PrPu}, boosted
\cite{boost1} and some preliminary unpublished results we have for
spinning holes. This would suggest that, if the same behavior takes
place for the inspiralling collisions,  ``reality'' should lie below
the curve corresponding to the Regge--Wheeler--Zerilli formalism
(Z). This would indicate that the estimation obtained using the Teukolsky
formalism is actually {\em worse} for the particular kind of collision
under consideration. This is what we were alluding to when we warned
in the introduction that it was not obvious that representing the
spacetime as a perturbation of a non-rotating hole was a worse choice
than of that of a Kerr hole.

\begin{figure}
\centerline{\psfig{figure=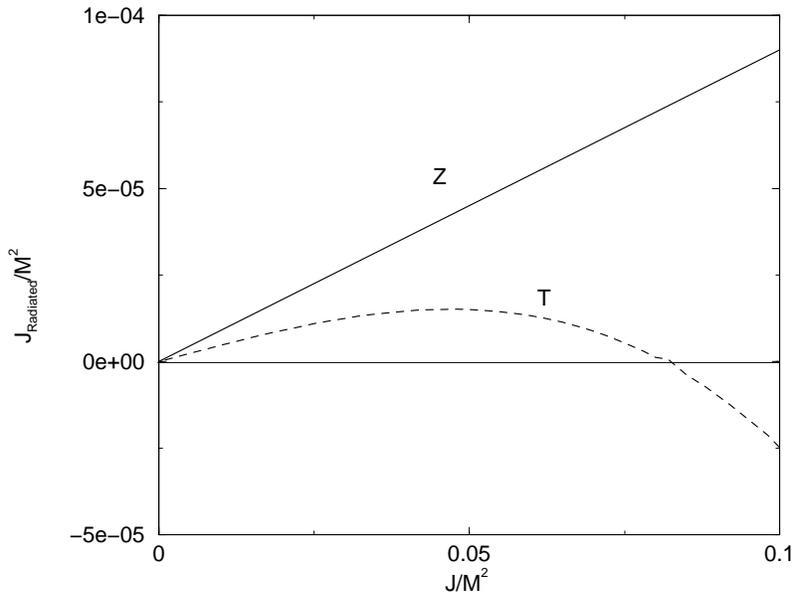,height=80mm}}
\caption{Radiation of angular momentum. As can be seen, the
Regge--Wheeler-Zerilli (Z) calculation of perturbations of a
non-rotating hole disagrees with the Teukolsky (T) rotating black hole
calculation. The radiated angular momentum is a more delicate quantity
to compute than the energy and it appears that the potentially
inconsistent higher order terms included in the rotating perturbation
approach changes its value significantly. Overall we see the radiation
is small. The Regge--Wheeler--Zerilli curve predicts less than $0.1\%$
of the total angular momentum will be radiated, even in the extreme
rotating case.}
\label{fig5}
\end{figure}

We now turn to the evaluation of the radiated angular momentum. This
is depicted in figure \ref{fig5}. 
The two curves shown in \ref{fig5} disagree significantly. They do not
even agree for very small values of the angular momentum. We have
checked that there is no numerical error: if in the Teukolsky
evolution one keeps the initial data intact but ``turns off'' the
$a$-dependent terms in the evolution equation, the RWZ straight line
is reproduced. It should be noticed that the radiated angular momentum
is a qualitatively different quantity insofar as its computation than
the energy. The energy is roughly obtained by squaring and integrating
the waveforms. The angular momentum depends on subtle phase
differences. It is much more easy to disturb the calculation of the
radiated angular momentum than that of the radiated energy. This, in
particular, points out to the potential difficulty of estimating this
kind of quantity in full numerical simulations, where phase lags in
the waveforms due to grid stretching and other problems are well
known. In our approach it appears that the potentially inconsistent
higher order terms in the angular momentum we introduce when
considering a rotating background are causing problems in the
computation of radiated angular momentum. If one wishes to be
ultra-conservative, one could simply conclude that both calculations
only predict the correct result for zero angular momentum. Otherwise,
one could conclude that for this family of initial data the Teukolsky
approach really only works for non-rotating black holes, something
suggested by the fact that the background spacetime is only recovered
in the close limit with vanishing angular momentum. At the moment we
can only say that the accurate computation of the radiated angular
momentum for this problem is an open problem. It is likely that the
RWZ estimate is correct, but we do not have ``error bars'' (even rough
ones) to validate this prediction.

\section{Conclusions}
We have used the ``close limit'' to estimate the radiation in the
collision at the end of the inspiral of two equal mass nonrotating
black holes. The assumptions and restrictions were: (i) only the
``ringdown'' radiation was computed; (ii) we assumed that a simple
initial data set gave an adequate representation of appropriate
astrophysical conditions; (iii) we assumed that the final hole is not
near the extreme Kerr limit; (iv) we used close limit estimates of the
evolution.  Our main conclusion is that the energy radiated in
ringdown is probably not more than 1\% of the total mass of the
system, and the angular momentum radiated is not more than 0.1\% of
the initial angular momentum. The most serious uncertainty in this
result is the possibility that the radiation from the early merger
stage of coalescence is very much larger than the ringdown radiation.
With our 1\%$Mc^{2}$ estimate, collisions of black holes of
$100M_{\odot}$ would be detectable with signal to noise of 6 out to
distances on the order of 200Mpc by the initial LIGO configuration and
to distances of 4Gpc with the advanced LIGO detector.

\section{Acknowledgements}
We wish to thank two anonymous referees for many constructive 
criticisms on the initial submitted version of the paper.
This work was supported in part by grants
NSF-INT-9512894,
NSF-PHY-9357219,
NSF-PHY-9423950,
NSF-PHY-9734871,
NSF-PHY-9800973,
NSF-PHY-9407194,
by funds of the University of
C\'ordoba, Utah, and Penn State.  We also acknowledge support of CONICET and
CONICOR (Argentina).  JP acknowledges support from the 
the John S.  Guggenheim foundation and hospitality from ITP at
UC Santa Barbara during completion of the manuscript.  
RJG is a member of CONICET.

\end{document}